\documentclass[twocolumn,secnumarabic,amssymb, amsmath, nofootinbib,tightenlines,
nobibnotes, aps, prb, showpacs]{revtex4}

\usepackage{graphicx}%
\usepackage{dcolumn}
\usepackage{amsmath}

\begin{document}

\preprint{HEP/123-qed}

\title[Short Title]{ 
Mass-Enhanced Fermi Liquid Ground State in Na$_{1.5}$Co$_2$O$_4$ 
}

\author{K. Miyoshi,$^1$ E. Morikuni,$^1$ K. Fujiwara,$^1$ J. Takeuchi$^1$, and T. Hamasaki$^2$}

\affiliation{%
$^1$Department of Material Science, Shimane University, Matsue 690-8504, Japan
}%
\affiliation{%
$^2$Department of Physics, Kyushu Sangyo University, Fukuoka 813-8503, Japan
}%

\date{\today}

\begin{abstract}

Magnetic, transport, and specific heat measurements 
have been performed on layered metallic oxide Na$_{1.5}$Co$_2$O$_4$ 
as a function of temperature $T$. 
Below a characteristic temperature $T^*$=30$-$40 K, 
electrical resistivity shows a metallic conductivity with a $T^2$ behavior 
and magnetic susceptibility deviates from the 
Curie-Weiss behavior showing a broad peak at $\sim$14 K. 
The electronic specific heat coefficient $\gamma$ is $\sim$60 mJ/molK$^2$ at 2 K. 
No evidence for magnetic ordering is found. 
These behaviors suggest the formation of mass-enhanced Fermi liquid ground state
analogous to that in $d$-electron heavy fermion compound LiV$_2$O$_4$.

\end{abstract}

\maketitle


There has been a great deal of interest in the physics 
of geometrically frustrated spin systems, 
in which 
the long-range magnetic order tends to be suppressed and 
novel types of phase transitions or ground states
are expected. 
In particular, the recent discovery of the heavy fermion (HF) behavior in 
LiV$_2$O$_4$ has stimulated 
new interest in the geometrically frustrated systems, 
where the HF behavior
is demonstrated by the $d$-electrons on the spinel B lattice.\cite{kondo_prl} 
One of the most exciting scenario is 
the generation of the $d$-electron heavy fermions  
through the effect coming from the geometrical frustration 
which is possibly inherent due to the geometry of V atoms 
and the negative Weiss temperature of $-$40 K.\cite{kondo_prl} 

The low-$T$ properties of LiV$_2$O$_4$ are 
apparently of intermetallic dense Kondo systems, showing a fairly large electronic specific heat 
coefficient $\gamma$$\sim$420 mJ/molK$^2$,\cite{kondo_prl,johnston} a broad maximum of 
magnetic susceptibility,\cite{kondo_prl,kondo_prb} and a metallic conductivity 
with a $T^2$ behavior below a coherence temperature 
$T^*$=20$-$30 K.\cite{urano}
Indeed, the origin of the HF behavior has been considered from various angles, 
but a uniform understanding remains to be achieved.  
Urano $et$ $al$. suggest the importance of the geometrical frustration,\cite{urano} which is 
inferred from the occurrence of spin-glass order 
by the slight substitution for the Li or V sites.\cite{ueda,onoda,miyoshi} 
A recent inelastic neutron scattering study has revealed a feature 
of frustrated magnetism in LiV$_2$O$_4$.\cite{lee} 
On the other hand, a broad maximum of $^7$Li $1/T_1$($T$) at 30$-$50 K and 
a constant $(T_{1}T)^{-1}$ at low $T$ have been found, 
suggesting a dense Kondo picture.\cite{mahajan} 
$^7$Li $1/T_1$($T$) has also been interpreted from the view point of 
the ferromagnetic instability predicted in the spin fluctuation theory.\cite{fujiwara98} 
From the theoretical side, 
various approaches have been introduced and 
several groups focus on 
the geometrical frustration.\cite{lacroix,burdin,hopkinson}

To gain more insight into this problem, 
the discovery of other examples of $d$-electron material to exhibit HF behavior 
is of significant importance.  
From this view point, it is interesting to note 
the low-$T$ properties of the metallic oxide 
$\gamma$-Na$_{x}$Co$_2$O$_4$ (1$\leq$$x$$\leq$1.5), 
which is well-known 
as a large thermoelectric material.\cite{terasaki} 
This compound has been found 
to exhibit a large $\gamma$ of $\sim$80 mJ/molK$^2$ at 2 K
and a Curie-Weiss behavior of magnetic susceptibility with a Weiss temperature 
$\Theta$$\sim$$-$120 K showing no sign of magnetic order.\cite{ando,ray,tojo} 
Na$_{x}$Co$_2$O$_4$ has a layered structure consisting of 
CoO$_2$ layers in which the Co atoms form a two-dimensional 
regular triangular lattice with interlayers of Na atoms alternatively 
stacked along the $c$ axis.\cite{fouassier} 
In addition to the large $\gamma$ value and the metallic conductivity, 
the absence of magnetic order and the arrangement of the magnetic ions sitting on a 
geometrically frustrated lattice 
are characteristic features in common with LiV$_2$O$_4$. 
Also, the C15 Laves phase compound (Y$_{0.95}$Sc$_{0.05}$)Mn$_2$, 
in which the arrangement of the Mn atoms is equivalent to B sites on spinel lattice, 
has been found to show no magnetic order and 
HF behavior with $\gamma$=150 mJ/molK$^2$.\cite{siga} 

In the present work, 
the intrinsic low-$T$ properties of Na$_{1.5}$Co$_2$O$_4$ 
have been investigated through the 
measurements of dc magnetic susceptibility ($\chi$), electrical 
resistivity ($\rho$) and specific heat ($C$). 
For the precise investigations, 
high purity specimens which were 
melted and grown in a floating-zone furnace 
were prepared. 
The crystal growth was performed 
in a high pressure oxygen atmosphere of 0.5$-$1.0 MPa.
The final products were not single crystals but multiple crystals. 
All the peaks observed in the powder X-ray diffraction 
were indexed as $\gamma$-Na$_{x}$Co$_2$O$_4$. 
The lattice constants  
were estimated to be $a$=2.84(5) \AA \ 
and $c$=10.85(2) \AA \ similar to those reported in the previous work.\cite{tanaka}
The chemical composition of Na$_{x}$Co$_2$O$_4$ was determined to be $x$$\sim$1.5 
by inductively coupled plasma analysis. 
dc magnetic susceptibility was measured by a superconducting quantum 
interference device (SQUID) magnetometer. 
Specific heat was measured by a thermal relaxation method. 
Electrical resistivity was measured using a standard 
four-probe technique.  

Figure 1 shows the $\chi$($T$) data for the melt-grown crystal 
of Na$_{1.5}$Co$_2$O$_4$ 
in a magnetic field of $H$=1 T.
A remarkable feature is a broad peak of $\chi$($T$) at 10$-$20 K, 
which is similar to that observed 
in LiV$_2$O$_4$.\cite{kondo_prl} 
No indication of magnetic order above 2 K was found in our 
field-cooled and zero-field-cooled $\chi$($T$) measurements. 
While for the sintered sample of Na$_{1.5}$Co$_2$O$_4$, 
the $\chi$($T$) curve has been reported to show a monotonic increase 
as decreasing temperature.\cite{tojo}
This is probably because the low-$T$ peak 
is masked by a Curie-like behavior of magnetic impurities or defects
contained in the sintered sample.  
Also for LiV$_2$O$_4$, a broad peak of $\chi$($T$) is 
often masked by a Curie-like behavior depending on the sample quality.\cite{kondo_prb} 
The $\chi$($T$) curve for the melt-grown crystal was fitted by the expression,
$\chi(T)$=$\chi_{0}$+$C$$/$($T$$-$$\Theta$) for 50$\leq$$T$$\leq$300 K, 
yielding a Curie constant $C$=0.369 emu/molK, a Weiss temperature 
$\Theta$=$-$139 K and a $T$-independent 
susceptibility $\chi_{0}$=1.12$\times$10$^{-4}$ emu/mol. 
For NaCo$_2$O$_4$, 
Co ion has a mixed valence between 
Co$^{3+}$ (3$d$$^6$) and Co$^{4+}$ (3$d$$^5$) (Co$^{3+}$/Co$^{4+}$=1), 
which are presumably in the 
low spin state with $S$=0 and $S$=1/2, respectively. 
The value of $C$ corresponds to an effective magnetic moment 
$\mu_{\rm eff}$$\sim$1.22 $\mu_{\rm B}$ per Co site. 

\begin{figure}[bp]
\includegraphics[width=8cm]{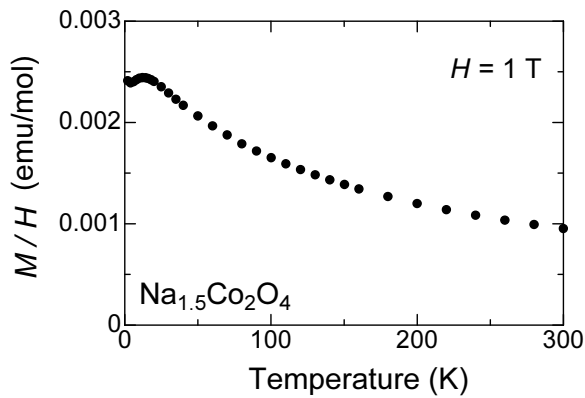}
\caption{
Plots of magnetic susceptibility vs temperature data 
for Na$_{1.5}$Co$_2$O$_4$ at $H$=1 T. 
}
\label{autonum}
\end{figure}

Next, we show the $\chi$($T$) data for Na$_{1.5}$Co$_2$O$_4$ 
at various magnetic fields up to 7 T in Fig. 2.  
The inset of Fig. 2 shows the $\chi$($T$) data for 
LiV$_2$O$_4$, which were collected using the specimen 
used in the previous work.\cite{miyoshi} 
In the $\chi$($T$) curves 
both for Na$_{1.5}$Co$_2$O$_4$ and LiV$_2$O$_4$, 
a broad peak is observed at 10$-$20 K, and a low-$T$ upturn seen at $H$=0.1 T 
vanishes at $H$=7 T. 
This is due to the saturation of the moments of the magnetic 
impurities in the high magnetic field, 
indicating that the intrinsic $\chi$($T$) behavior is observed at $H$=7 T.   
A remarkable feature is that 
the height and position of the peak commonly appear to 
be almost unchanged by the application of high magnetic field. 
Moreover, the peak temperature of $\chi$($T$) at $H$=7 T for Na$_{1.5}$Co$_2$O$_4$ 
is identified as $T_{\rm p}$$\sim$14 K, 
similar to that for LiV$_2$O$_4$ ($T_{\rm p}$=15$-$16 K).  
The broad peak behavior in $\chi$($T$) 
is similarly observed in intermetallic dense Kondo systems, 
which are thought to become Pauli paramagnetic
below the peak temperature. 
The broad peak behavior, however, 
does not agree with $S$=1/2 Kondo model predictions,\cite{kondo_prb} 
since the behavior is limited to the model with $J$$\geq$3/2.\cite{rajan} 

\begin{figure}[bp]
\includegraphics[width=8cm]{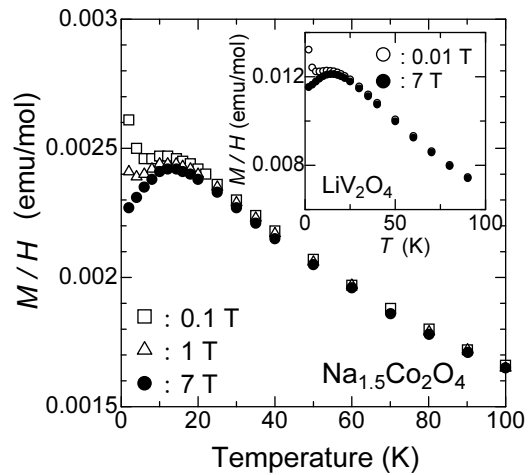}
\caption{
Plots of magnetic susceptibility vs temperature data 
for the melt-grown crystal of Na$_{1.5}$Co$_2$O$_4$ in 
various magnetic fields up to 7 T. 
The inset shows those for LiV$_2$O$_4$. 
}
\label{autonum}
\end{figure}

For LiV$_2$O$_4$, 
a metallic behavior with 
$\rho$($T$)$\propto$$T^2$ has been observed 
below $T^{*}$=20$-$30 K, 
indicating the formation of heavy-mass Fermi liquid (FL) below $T^{*}$.\cite{urano} 
The results of the $\rho$($T$) measurements for the melt-grown crystal 
of Na$_{1.5}$Co$_2$O$_4$ are displayed in Fig. 3. 
$\rho$($T$) was measured along the cleaved (001) surface of the melt-grown crystal. 
The sample was not a single crystal but we think that the data 
roughly represent the in-plane resistivity due to the crystal orientation. 
In Fig. 3, the $\rho$($T$) curve at ambient pressure 
decreases as the temperature is lowered below 300 K and 
shows a higher rate of decrease below 30$-$40 K. 
The inset of Fig. 3 shows a plot of $\rho$ versus $T^2$ 
at low $T$, clearly indicating a $T^2$ behavior in $\rho$($T$). 
Thus, the $\rho$($T$) data suggest the FL behavior below $T^{*}$=30$-$40 K. 
$T^{*}$ was found to nearly correspond to the temperature 
below which the deviation from the 
Curie-Weiss behavior in $\chi$($T$) is apparent, as in LiV$_2$O$_4$.\cite{urano} 
We obtain a coefficient of the $T^2$ term $A$=0.48 $\mu$$\Omega$cm$/$K$^{2}$, 
which yields 
$A/\gamma^2$$\sim$1.3$\times$10$^{-4}$ $\mu$$\Omega$cm$/$K$^{2}$$/$(mJ$/$mol K$^{-2}$)$^{2}$. 
The value of $A/\gamma^2$ is fairly larger than 
that for the Kadowaki-Woods relation, 
i.e., $A/\gamma^2$$\sim$10$^{-5}$ 
$\mu$$\Omega$cm$/$K$^{2}$$/$(mJ$/$mol K$^{-2}$)$^{2}$, 
but the value would be reduced 
if we used a single crystal specimen for the $\rho$($T$) measurement. 

The $\rho$($T$) data in Fig. 3 is inconsistent with 
that measured using a single crystal specimen of NaCo$_2$O$_4$ 
prepared by NaCl-flux technique in the previous work,\cite{terasaki} 
where $\rho$($T$) along the (001) plane decreases monotonously as decreasing temperature even at low $T$. 
The origin of the discrepancy is unclear but 
a similar discrepancy has also been observed between sintered samples,\cite{kawata,motohashi}
one of which shows a metallic behavior 
with a marked decrease in $\rho$($T$) below $\sim$40 K similar to 
$\rho$($T$) in Fig. 3 and is prepared by ``rapid heat-up technique''.\cite{motohashi} 
One may consider that the $\rho$($T$) data in Fig. 3 
reflects not only in-plane resistivity but also out-of-plane resistivity, 
which has been reported to show a maximum at $\sim$200 K.\cite{terasaki} 
However, we note that $\rho$($T$) in Fig. 3 
shows no tendency to make a maximum at $\sim$200 K and out-of-plane resistivity 
reported previously does not show a marked decrease below 40 K.\cite{terasaki}

\begin{figure}[bp]
\includegraphics[width=8cm]{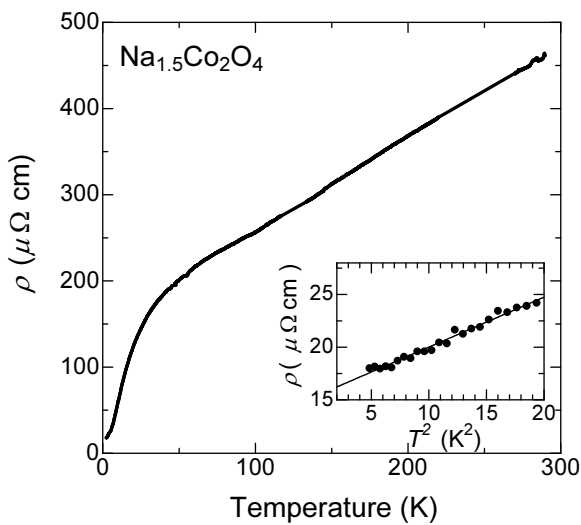}
\caption{
Temperature dependence of electrical resistivity 
for the melt-grown crystal of Na$_{1.5}$Co$_2$O$_4$. 
The inset shows the data of electrical resistivity 
plotted against $T^2$. 
}
\label{autonum}
\end{figure}

Next, we investigate the behavior of $C/T$ vs $T$ 
for the melt-grown crystal of Na$_{1.5}$Co$_2$O$_4$. 
The result is shown in Fig. 4, 
where $C/T$ shows a gradual decrease with decreasing temperature. 
The behavior is inconsistent with that for 
the sintered sample of NaCo$_2$O$_4$ in the previous work, 
where $C/T$ increases with decreasing temperature below 5 K 
and reaches 80 mJ/molK$^2$ at 2 K.\cite{ando} 
For a sintered sample, 
the contribution of the magnetic impurities, 
associated with the release of the magnetic entropy $S$=$\int$$C/TdT$ at low $T$, 
could not be excluded. 
In Fig. 4, $\gamma$(2 K) ($\equiv$ $C/T$ at 2 K) for Na$_{1.5}$Co$_2$O$_4$ 
is estimated to be $\sim$60 mJ/molK$^2$, 
which is three times larger than the value obtained from 
a recent band calculation.\cite{singh} 
The $C/T$$-$$T$ curve is qualitatively similar to that for 
UGa$_3$ ($\gamma$=43 mJ/molK$^2$).\cite{cornelius} 
Using $\chi$(2 K)=22 mJ/molT$^2$ and $\gamma$(2 K)=60 mJ/molK$^2$, 
we obtain a Wilson ratio $R_{\rm w}$$\sim$2.7. 
No indication of magnetic order 
was found for both samples in the measurements for 2$\leq$$T$$\leq$100 K.

\begin{figure}[bp]
\includegraphics[width=8cm]{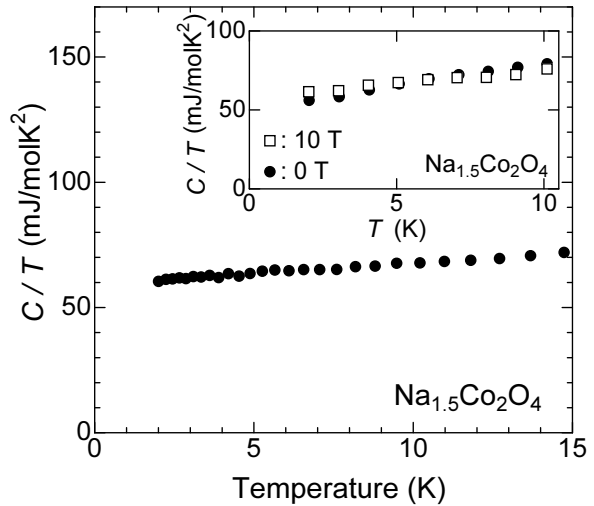}
\caption{
Plots of specific heat divided by temperature $C/T$ vs temperature $T$ 
for the melt-grown crystal of Na$_{1.5}$Co$_2$O$_4$. 
The inset shows 
plots of $C/T$ vs $T$ in a magnetic field of 10 T (open squares) 
and in zero field (closed circles). 
}
\label{autonum}
\end{figure}

For LiV$_2$O$_4$, $\gamma$($T$) is fitted by the prediction 
for the $S$=$1/2$ Kondo model, yielding 
a Kondo temperature $T_{\rm K}$=27.5 K.\cite{kondo_prl} 
Assuming one spin ($S$=$1/2$) per formula unit, 
the $C/T$$-$$T$ curve for the melt-grown crystal of Na$_{1.5}$Co$_2$O$_4$ was also 
reproduced by the prediction for 
an $S$=$1/2$ Kondo model (Ref. 26) with $T_{\rm K}$=140 K 
at least in the limited temperature range 2$\le$$T$$\le$15 K.
However, $T_{\rm K}$ is fairly higher than 
$T_{\rm p}$ ($\sim$14 K) for $\chi$($T$) and 
$T^{\rm *}$ (=30$-$40 K) for $\rho$($T$), 
inconsistent with a dense Kondo picture.
In the inset of Fig. 4, the $C/T$$-$$T$ curve 
for the melt-grown crystal of Na$_{1.5}$Co$_2$O$_4$ 
in a magnetic field of $H$=10 T, together with the zero-field data, is shown. 
$\gamma$(2 K) for $H$=10 T was estimated to be about 10$\%$ 
larger than that for zero field. 
This is contrary to that expected in conventional dense Kondo systems, 
where the value of $\gamma$ tends to decrease with applied magnetic field 
and the field dependence is explained by the broadening of 
the Kondo resonance with applied field.\cite{andraka,satoh} 
Na$_{1.5}$Co$_2$O$_4$ is unlike dense Kondo systems. 

It should be noted that, in both of Na$_{1.5}$Co$_2$O$_4$ and LiV$_2$O$_4$, 
below $T^{\rm *}$, 
the system shows FL behavior accompanied by a large mass enhancement 
and $\chi(T)$ deviates from the Curie-Weiss behavior showing a 
broad peak at 14$-$16 K. 
These behaviors suggest that the mass-enhanced FL 
ground states in these compounds are analogous to each other. 
A Kondo lattice model has been proposed for LiV$_2$O$_4$,\cite{anisimov} but 
is not adequate as a unified description for these compounds, 
because Na$_{1.5}$Co$_2$O$_4$ is unlike dense Kondo systems as mentioned above. 
Moreover, it is not obvious how the localized moments and itinerant carriers 
as in $f$-electron HF compounds originate from 
the same 3$d$ shell with the electron configuration 3$d^{\rm 5.5}$ at low spin state.  
A common feature between Na$_{1.5}$Co$_2$O$_4$ and LiV$_2$O$_4$ is 
the geometry of magnetic ions possibly giving rise to the spin frustration. 
Furthermore, HF behavior has been found in 
(Y$_{0.95}$Sc$_{0.05}$)Mn$_2$ ($\gamma$=150 mJ/molK$^2$) (Ref. 19) 
and $\beta$-Mn ($\gamma$=70 mJ/molK$^2$),\cite{shinkoda} both of which have 
magnetic ions sitting on a geometrically frustrated lattice and do not 
show any magnetic order. 
It is hard to imagine that the common structural feature possibly leading to 
the spin frustration in the mass-enhanced FL 
compounds noted above is accidental. 

In summary, 
our measurements 
for the melt-grown crystal of Na$_{1.5}$Co$_2$O$_4$ 
have revealed a large $\gamma$ value of $\sim$60 mJ/molK$^2$ at 2 K, 
a $T^2$ behavior of $\rho(T)$ 
and a characteristic broad peak behavior of $\chi(T)$ at $\sim$14 K, 
suggesting the formation of mass-enhanced FL analogous 
to that in LiV$_2$O$_4$ below $T^{*}$=30$-$40 K. 
Absence of magnetic order and mass-enhanced FL behavior 
are common features among Na$_{1.5}$Co$_2$O$_4$, LiV$_2$O$_4$, 
(Y$_{0.95}$Sc$_{0.05}$)Mn$_2$ and $\beta$-Mn, all of which have a magnetic sublattice 
identical to that in geometrically frustrated systems.       
To advance towards further understanding for what happens in these systems, 
the theoretical explanation is desirable for 
the origin of the broad peak behavior in $\chi(T)$ 
and the role of the geometrically frustrated lattice in the ground state.

This work is supported in part by a Grant-in-Aid for 
Scientific Research (No. 15740215) from the Japanese 
Ministry of Education, Culture, Sports, Science and Technology.


\begin{thebibliography}{99}

\bibitem{kondo_prl} S. Kondo, D. C. Johnston, C. A. Swenson, F. Borsa, A. V. Mahajan, 
L. L. Miller, T. Gu, A. I. Goldman, 
M. B. Maple, D. A. Gajewski, E. J. Freeman, N. R. Dilley, R. P. Dickey, 
J. Merrin, K. Kojima, G. M. Luke, Y. J. Uemura, O. Chmaissem, and J. D. Jorgensen, 
Phys. Rev. Lett. {\bf  78}, 3729 (1997). 

\bibitem{johnston} D. C. Johnston, C. A. Swenson, and S. Kondo, 
Phys. Rev. B {\bf 59}, 2627 (1999). 

\bibitem{kondo_prb} S. Kondo, D. C. Johnston, and L. L. Miller, 
Phys. Rev. B {\bf 59}, 2609 (1999).

\bibitem{urano}  C. Urano, M. Nohara, S. Kondo, F. Sakai, H. Takagi, 
T. Shiraki, and T. Okubo, 
Phys. Rev. Lett. {\bf 85}, 1052 (2000).

\bibitem{ueda} Y. Ueda, N. Fujiwara, and H. Yasuoka, 
J. Phys. Soc. Japan {\bf 66}, 778 (1997).

\bibitem{onoda} M. Onoda, H. Imai, Y. Amako, and H. Nagasawa, 
Phys. Rev. B {\bf 56}, 3760 (1997).

\bibitem{miyoshi} K. Miyoshi, M. Ihara, K. Fujiwara, and J. Takeuchi, 
Phys. Rev. B {\bf 65}, 92414 (2002). 

\bibitem{lee} S. -H. Lee, Y. Qiu, C. Broholm, Y. Ueda, and J. J. Rush, 
Phys. Rev. Lett. {\bf 86}, 5554 (2001).

\bibitem{mahajan} A. V. Mahajan, R. Sala, E. Lee, F. Borsa, S. Kondo, and D. C. Johnston, 
Phys. Rev. B {\bf 57}, 8890 (1998). 

\bibitem{fujiwara98} N. Fujiwara, H. Yasuoka, and Y. Ueda, 
Phys. Rev. B {\bf 57}, 3539 (1998).

\bibitem{lacroix} C. Lacroix, Can. J. Phys. {\bf 79}, 1469 (2001).

\bibitem{burdin} S. Burdin, D. R. Grempel, and A. Georges, 
Phys. Rev. B {\bf 66}, 45111 (2002).

\bibitem{hopkinson} J. Hopkinson and P. Coleman, 
Phys. Rev. Lett. {\bf 89}, 267201 (2002). 

\bibitem{terasaki} I. Terasaki, Y. Sasago, and K. Uchinokura, 
Phys. Rev. B {\bf 56}, R12685 (1997).

\bibitem{ando} Y. Ando, N. Miyamoto, K. Segawa, T. Kawata, and I. Terasaki, 
Phys. Rev. B {\bf 60}, 10580 (1999). 

\bibitem{ray} R. Ray, A. Ghoshray, K. Ghoshray, and S. Nakamura, 
Phys. Rev. B {\bf 59}, 9454 (1999). 

\bibitem{tojo} T. Tojo, H. Kawaji, T. Atake, Y. Yamamura, M. Hashida, and T. Tsuji, 
Phys. Rev. B {\bf 65}, 52105 (2002). 

\bibitem{fouassier} C. Fouassier, G. Matejka, J. -M. Reau, and P. Hagenmuller, 
J. Solid State Chem. {\bf 6}, 532 (1973). 

\bibitem{siga} M. Shiga, K. Fujisawa and H. Wada, 
J. Phys. Soc. Japan {\bf 62}, 1329 (1993).

\bibitem{tanaka} T. Tanaka, S. Nakamura, and S. Iida, 
Jpn. J. Appl. Phys. {\bf 33}, L581 (1994). 

\bibitem{rajan} V. T. Rajan, 
Phys. Rev. Lett. {\bf 51}, 308 (1983).

\bibitem{kawata} T. Kawata, Y. Iguchi, T. Itoh, K. Takahata, and I. Terasaki, 
Phys. Rev. B {\bf 60}, 10584 (1999). 

\bibitem{motohashi} T. Motohashi, E. Naujalis, R. Ueda, K. Isawa, 
M. Karppinen, and H. Yamauchi, 
Appl. Phys. Lett. {\bf 79}, 1480 (2001). 

\bibitem{singh} D. J. Singh, 
Phys. Rev. B {\bf 61}, 13397 (2000). 

\bibitem{cornelius} A. L. Cornelius, A. J. Arko, J. L. Sarrao, J. D. Thompson, 
M. F. Hundley, C. H. Booth, N. Harrison, and P. M. Oppeneer, 
Phys. Rev. B {\bf 59}, 14473 (1999). 

\bibitem{desgranges} H. -U. Desgranges and K. D. Schotte, Phys. Lett. {\bf 91A}, 240 (1982).  

\bibitem{satoh} K. Satoh, T. Fujita, Y. Maeno, 
Y. $\bar{\rm O}$nuki, T. Komatsubara, and T. Ohtsuka, 
Solid State Commun. {\bf 56}, 327 (1985). 

\bibitem{andraka} B. Andraka, J. S. Kim, G. R. Stewart, and Z. Fisk, 
Phys. Rev. B {\bf 44}, 4371 (1991). 

\bibitem{anisimov} V. I. Anisimov, M. A. Korotin, M. Z$\ddot{\rm o}$lfl, 
T. Pruschke, K. Le Hur, and T. M. Rice, 
Phys. Rev. Lett. {\bf 83}, 364 (1999). 

\bibitem{shinkoda} T. Shinkoda, K. Kumagai, and K. Asayama,
J. Phys. Soc. Japan {\bf 46}, 1754 (1979).

\end{thebibliography}
\end{document}